# Quantifying vitamin D intake among Aboriginal and Torres Strait Islander peoples in Australia

Belinda Neo [1], Dale Tilbrook [2], Noel Nannup [3], Alison Daly [4], Eleanor Dunlop [4,5], John Jacky [3], Carol Michie [3], Cindy Prior [3], Brad Farrant [3], Carrington C.J. Shepherd [1,3,6], and Lucinda J. Black [4,5],*

[1] Curtin Medical School, Curtin University, Bentley, Western Australia, Australia
[2] Maalingup Aboriginal Gallery, Caversham, Western Australia, Australia
[3] Telethon Kids Institute, The University of Western Australia, Nedlands, Western Australia, Australia
[4] Curtin School of Population Health, Curtin University, Bentley, Western Australia, Australia
[5] Institute for Physical Activity and Nutrition (IPAN), School of Exercise and Nutrition Sciences, Deakin University, Geelong, Victoria, Australia
[6] Ngangk Yira Institute, Murdoch University, Murdoch, Western Australia, Australia

\* Correspondence: Lucinda J. Black and Lucinda.black@deakin.edu.au

**Abstract**

**Background/Objective:** Vitamin D deficiency (serum 25-hydroxyvitamin D [25(OH)D] concentration <50 nmol/L) is prevalent among Aboriginal and Torres Strait Islander peoples in Australia. Alternative to sun exposure (the primary source of vitamin D), vitamin D can also be obtained from food (e.g., fish, eggs, and meat) and supplements. However, vitamin D intake among Aboriginal and Torres Strait Islander peoples is currently unknown. We aimed to provide the first quantification of vitamin D intake using nationally representative data from Aboriginal and Torres Strait Islander peoples.

**Methods:** We used food consumption data collected in the 2012-2013 National Aboriginal and Torres Strait Islander Nutrition and Physical Activity Survey (n = 4,109) and vitamin D food composition data to quantify mean absolute vitamin D intake by sex, age group, and remoteness of location. Differences in mean vitamin D intake between sexes and between remoteness of location were assessed using the 95% confidence interval (95% CI).

**Results:** The mean (standard deviation (SD)) vitamin D intake among Aboriginal and Torres Strait Islander peoples was 2.9 (3.0) µg/day. Males had a statistically significantly higher mean (SD) [95% CI] vitamin D intake (3.2 (3.1) [3.0-3.4] µg/day) than females (2.6 (2.7) [2.4-2.7] µg/day). There were no statistically significant differences between mean (SD) [95% CI] vitamin D intake in non-remote (2.9 (2.2) [2.7-3.1] µg/day) and remote areas (2.8 (4.8) [2.6-3.0] µg/day).

**Conclusions:** Vitamin D intake among Aboriginal and Torres Strait Islander peoples is low. Food-based public health strategies could be developed to promote higher vitamin D intake among this population.

**Introduction**

Vitamin D deficiency (serum 25-hydroxyvitamin D [25(OH)D] concentration <50 nmol/L) is prevalent among Aboriginal and Torres Strait Islander peoples (27% of adults aged ≥18 years, increasing to 39% in those living remotely in Australia) (1). Vitamin D deficiency may result in adverse health outcomes in musculoskeletal health (2) and non-communicable diseases (e.g., type 2 diabetes and cardiovascular diseases) which are highly prevalent among Aboriginal and Torres Strait Islander peoples (3, 4).

The primary source of vitamin D is ultraviolet-B radiation from sun exposure. When sun exposure is minimal, dietary vitamin D intake is crucial to prevent vitamin D deficiency. The estimated average requirement for vitamin D intake recommended by the Institute of Medicine (assuming minimal sunlight exposure) is 10 μg/day for people aged ≥ 1 years (5). Recently, four D vitamers were measured in retail food products and game products to develop Australia's first comprehensive national vitamin D food composition database (6, 7). Using those composition data, we previously quantified vitamin D intake in the general Australian population (8); however, vitamin D intake among Aboriginal and Torres Strait Islander peoples remains unquantified.

The 2012-2013 National Aboriginal and Torres Strait Islander Nutrition and Physical Activity Survey (NATSINPAS) was the first national survey to collect detailed national benchmark nutrition information among Aboriginal and Torres Strait Islander peoples. The NATSINPAS, in combination with the existing comprehensive vitamin D composition data, enabled us to quantify the first mean absolute vitamin D using nationally representative data from Aboriginal and Torres Strait Islander peoples.

**Methods**

**Study population**

The NATSINPAS was conducted between August 2012 and July 2013 and provided the most recent food consumption data among Aboriginal and Torres Strait Islander peoples in Australia. Detailed methods can be found elsewhere (9). Briefly, the survey was conducted across all states and territories in Australia and included participants from 2,900 private dwellings. The remoteness of the location of residence was divided into two categories, remote and non-remote, based on the Australian Statistical Geography Standard classification (9). The total number of participants was 4,109, with 1,792 living in non-remote areas and 2,317 living in remote areas.

The first survey interview was conducted face-to-face by a trained interviewer. An adult household member (aged ≥18 years) responded on behalf of those aged <15 years, and on behalf of those aged 15-17 years who did not have parental consent to respond independently. Participants living in non-remote areas were invited to complete a second interview via telephone eight days after the first interview.

The response rate for the first interview was 99.5%, and missing data for the remaining 0.5% were imputed by the Australian Bureau of Statistics (10). The second interview was only conducted among participants living in non-remote areas, and the overall response rate was low (n = 771). Hence, we were unable to use day 2 intake data to estimate usual vitamin D intakes.

**Food consumption data**

A 24-hour dietary food recall was conducted in the 2012-2013 NATSINPAS to collect detailed food consumption data of Aboriginal and Torres Strait Islander peoples aged ≥2 years. Detailed methods are reported elsewhere (9) and briefly summarised here. For the 24-hour food recall interview segment, children aged 6-14 years were encouraged to participate with an adult household member present instead of engaging an interview proxy (10). For those aged 15-17 years, 61% were interviewed individually, and 39% were interviewed with an adult household member.

The Automated Multiple-Pass Method was developed by the United States Department of Agriculture (11). It was adapted for use in Australia by Food Standards Australia New Zealand and integrated into the Computer Assisted Personal Interview instrument (9). Participants were asked systematic questions based on the Automated Multiple-Pass Method prompts, designed to accurately capture detailed information about the food, cooking methods, and portion sizes they consumed from midnight to midnight on the day before the interview. The Australian Health Survey (AHS) food model booklet (12) and bush tucker prompt cards (13) were used to help participants estimate the amount of food and beverages consumed. Food consumption data were subsequently coded using the AHS classification system, an eight-digit numeric food identification code (14).

**Vitamin D composition data**

We published vitamin D composition data for four D vitamers (vitamin $D_3$, $25(OH)D_3$, vitamin $D_2$, and $25(OH)D_2$) in retail food products and game products with detailed methods (6, 7). Game products form part of the traditional diets of Aboriginal and Torres Strait Islander peoples (15, 16). Briefly, to test for the presence of D vitamers, retail food products expected to contain vitamin D were sampled across three Australian cities (Sydney,

Melbourne, and Perth). Game meats, including camel, crocodile, emu, and kangaroo, were purchased Victoria, and emu eggs and oil were purchased from emu farms in Western Australia, Victoria, and New South Wales. The D vitamers in these foods were measured using liquid chromatography with triple quadrupole mass spectrometry at the National Measurement Institute of Australia, a laboratory accredited by the National Association of Testing Authorities to measure vitamin D in food (ISO17025:2017).

The Australian Food and Nutrient Database 2011-2013 is a food nutrient database used to estimate nutrient intake from the AHS (14). We previously mapped the analytical concentrations of D vitamers for each sampled food to the 5,740 Australian Food and Nutrient Database 2011-13 food entries (14), following the method used for the Australian total diet studies (17).

**Bioactivity of 25(OH)D**

Biological activity refers to the bioaccessibility (the amount of nutrient potentially available for absorption) and bioavailability (the amount of nutrient absorbed that is available to be used and stored) of the nutrient (18). A bioactivity factor of 1 assumes equal biological activity of vitamers. Among the four D vitamers, the hydroxylated D vitamers ($25(OH)D_2$ and $25(OH)D_3$) may have up to five times higher biological activity compared to vitamin $D_2$ and vitamin $D_3$ (19, 20). Hence, we quantified vitamin D intake using bioactivity factors 1 and 5 for 25(OH)D.

**Quantifying absolute vitamin D intake**

We quantified the mean (SD) for absolute vitamin D intake using day 1 of 24-hour food recall data from NATSINPAS. The weight of each food consumed by each respondent per day (g) was multiplied by the vitamin D content of the food (µg/g). Person weights from the Australian Bureau of Statistics (9) were used to quantify population vitamin D intake. Statistical analysis was performed using Stata version 17 (StataCorp, College Station, Texas, USA). The results were presented by sex, age group, and remoteness of location. We compared the difference in mean vitamin D intake between sex and between remoteness of location using the 95% confidence interval (95% CI) generated by the two-way table survey analysis: intakes were considered statistically significantly different if the 95% CI did not overlap.

**Percentage of food group contribution to vitamin D intake**

The percentage contribution of food groups according to the AHS classification system (14) to vitamin D intake was calculated as (total vitamin D intake from food group divided by total vitamin D intake from all foods) x 100 (21). We calculated the 95% CI using an exact binomial test. The top 10 food group contributors to total vitamin D intake were reported.

**Results**

**Bioactivity factor 1**

Across all ages for both sexes, the mean (SD) vitamin D intake was 2.9 (3.0) µg/day (**Table 1**). Children aged 2-3 years had the lowest mean vitamin D intake, and children aged 9-13 years had the highest mean vitamin D intake. The mean (SD) [95% CI] vitamin D intake in males (3.2 (3.1) [3.0-3.4] µg/day) was statistically significantly higher than in females (2.6 (2.7) [2.4-2.7] µg/day). Older female adults aged ≥ 71 years had the lowest mean (SD) vitamin D intake at 1.7 (2.9) µg/day, and male children aged 9-13 years had the highest mean (SD) vitamin D intake at 3.7 (4.3) µg/day.

There were no significant differences between mean (SD) [95% CI] vitamin D intake in non-remote (2.9 (2.2) [2.7-3.1] μg/day) and remote areas (2.8 (4.8) [2.6-3.0] μg/day) **(Table 2)**. Vitamin D intake for non-remote and remote areas by sex and age group are reported in **Supplementary Tables 1 and 2**, respectively.

Across the age groups and sexes, 'Meat, poultry, game products and dishes' had the highest contribution to total vitamin D intake, followed by 'Fats and oils,' and 'Egg products and dishes' **(Table 3)**. The top 10 food group contributors to total vitamin D intake for non-remote and remote areas are found in **Supplementary Tables 3 and 4**, respectively.

**Bioactivity factor 5**

When a bioactivity factor of 5 was applied, the mean (SD) vitamin D intake was 5.3 (4.3) μg/day, and the highest mean (SD) vitamin D intake was observed in adults aged 19-30 years at 5.9 (4.1) μg/day **(Table 1)**.

Similar to bioactivity factor 1, children aged 2-3 years had the lowest mean (SD) vitamin D intake at 3.5 (2.6) μg/day. The mean (SD) [95% CI] vitamin D intake was also significantly higher in males at 5.9 (4.4) [5.6-6.2] μg/day than in females at 4.7 (4.0) [4.4-4.9] μg/day. There were also no significant differences in mean (SD) [95% CI] vitamin D intake between non-remote (5.2 (3.2) [5.0-5.5] μg/day) and remote areas (5.4 (7.3) [5.2-5.7] μg/day). The top three food group contributors to total vitamin D intake were 'Meat, poultry, game products and dishes,' 'Egg products and dishes,' and 'Fats and oils' **(Table 3)**.

**Discussion**

Similar to the general Australian population (8), vitamin D intakes were low among Aboriginal and Torres Strait Islander peoples of all age groups and sexes for both bioactivity factors 1 and 5 for 25(OH)D. Our findings also indicate that males had a statistically significantly higher mean vitamin D intake compared to females, similar to our previous analysis in the general Australian population (8). The higher vitamin D intake among males may be due to the higher overall food intake compared to females (10). Given the high prevalence of vitamin D deficiency (1) and low vitamin D intakes among Aboriginal and Torres Strait Islander peoples, evidence-based public health strategies are needed to promote vitamin D sufficiency.

Assuming equal bioactivity among D vitamers, 'Meat, poultry, game products, and dishes' had the greatest contribution to vitamin D intake among Aboriginal and Torres Strait Islander peoples, indicating that those foods are widely consumed in this population. Comparatively, 'Fish and seafood products and dishes' contributed the greatest to vitamin D intake in the general Australian population (8). Given that edible oil spreads (e.g., margarine) are the only foods mandated for vitamin D fortification in Australia (22), it was not surprising that 'Fats and oils' was one of the top three food group contributors to vitamin D intake for both bioactivity factors 1 and 5 for 25(OH)D. Foods such as beef, chicken, and eggs have a higher concentration of 25(OH)D compared to other foods (e.g., margarine, fin fish with <5% fat, and seafood products) (7). When a bioactivity of 5 for 25(OH)D was applied, 'Meat, poultry, game products and dishes' was the greatest contributor to vitamin D intake for Aboriginal and Torres Strait Islander people and the general Australian population (8).

Vitamin D intake almost doubled across all sex and age groups when we applied a bioactivity factor of 5. Given the lack of consensus regarding the bioactivity of 25(OH)D, some international food composition databases have chosen to use a bioactivity factor of 5 (e.g., United Kingdom) (23), while others assume equal bioactivity across all D vitamers (e.g., Canada) (18, 24). We used bioactivity factors of both 1 and 5 for 25(OH)D, which allows for the comparison of vitamin D intake with other countries. However, as there is no current international consensus on the bioactivity factor of 25(OH)D, it has been suggested that equal bioactivity should be assumed until more definitive results are available (18).

The data on vitamin D intake among Indigenous peoples worldwide are notably scarce, which might be attributable to a lack of nationally representative food consumption data for those population groups (25, 26). Notable exceptions include subsets of populations such as the (i) Dené/Métis communities in Canada (mean vitamin D intake at 4.6 µg/day for those aged 7-85 years) (27); (ii) lactating Inuit women in the Inuvialuit Settlement Region, Nunavut, and Nunatsiavut communities in Canada (mean (SD) vitamin D intake at 4.8 (5.2) µg/day) aged > 18 years (28); (iii) Māori people in New Zealand (median (interquartile range) 3.2 (3.9) µg/day for those aged 80-90 years) (29); and (iv) Sami people in northern Norway (median vitamin D intake at 11.1 µg/day for males and 9.4 µg/day for females aged 40-69 years) (30). In line with our findings, vitamin D intake among the Indigenous peoples were predominantly low. Vitamin D intake quantified in these studies, however, might be underestimated as not all D vitamers were accounted for in the food composition data, especially 25(OH)D, which may have greater biological activity than vitamin D.

Low vitamin D intake among Indigenous peoples (27-30) could be due to the dietary transition of food preferences, where consumption of game products has declined (31). Traditional foods are considered valuable micronutrient sources such as vitamin D (25, 32). However, colonisation, modernisation, and food insecurity have altered the food availability and choices of Indigenous people worldwide (25, 33, 34). A similar transition has been seen in the diet of Aboriginal and Torres Strait Islander peoples since the colonisation of Australia around 250 years ago, where Aboriginal and Torres Strait Islander peoples were forcibly relocated and denied access to traditional foods and land (33). Traditional foods were replaced by food rations imposed through government policy, which were energy-dense, and high in sugar, sodium, and fat (33, 35). Hence, strategies to promote the cultivation and consumption of nutrient-rich traditional foods could be developed through co-designed research led by Elders and Aboriginal and Torres Strait Islander peoples, drawing upon their traditional knowledge of growing and preparing traditional foods. Elders are highly respected, influential individuals in their community with a wealth of traditional knowledge and experience in traditional foods (36). Promoting traditional ways of eating may improve vitamin D intake and status among Aboriginal and Torres Strait Islander peoples.

Given the prevalence of vitamin D deficiency in Australia, which affects 27% of Aboriginal and Torres Strait Islander peoples aged ≥ 18 years (1) and 20% of general Australians aged ≥ 25 years (37), there is a need to explore population-based dietary approaches such as supplementation and food fortification to promote vitamin D sufficiency. For example, countries such as Canada and Denmark have recommended vitamin D supplementation for specific at-risk population groups (e.g., older adults) (38, 39). However, potential limitations for implementing supplementation as a population-wide public health strategy include adherence and potential vitamin D toxicity (40). There is currently no public health recommendation for vitamin D supplementation in Australia.

Alternatively, most countries have used food fortification of commonly consumed foods (e.g., milk, bread, and breakfast cereals) as a public health strategy to increase vitamin D intake (41-43). For example, after vitamin D food fortification was implemented in Finland, mean vitamin D intake increased from 5 to 17 µg/day for males and 3 to 18 µg/day for females from 2002 to 2012 (41). In Australia, besides food mandated for vitamin D fortification (namely edible oil spreads e.g., margarine), very few eligible products are voluntarily fortified (44). We recently modelled vitamin D food fortification of fluid milk and alternatives using nationally-representative food consumption data and showed potential increases in vitamin D intake of about 2 µg/day among the general Australian population (44). Our findings will allow similar modelling of vitamin D fortification scenarios using commonly consumed staple foods such as bread, dairy milk, and breakfast cereals, consumed by 70%, 69%, and 34% of Aboriginal and Torres Strait Islander peoples, respectively (10). The modelling of vitamin D fortification could support potential policy and practice decisions to promote vitamin D sufficiency across the entire Australian population.

A strength of our study was using the most comprehensive nationally representative food consumption data and comprehensive food composition data for retail food products and game products in Australia. We quantified vitamin D intake using food composition data that comprised all four D vitamers, and we used both bioactivity factors 1 and 5 for 25(OH)D to account for the higher biological activity of 25(OH)D. Our findings provide a baseline to monitor trends in vitamin D intake over time in the Aboriginal and Torres Strait Islander population.

Our study was limited by the availability of only one day of 24-hour food recall. Modelling the usual nutrient intake of a population requires at least two days of 24-hour food recall data to eliminate the within-subject variation of dietary intake (45, 46). As 24-hour food recall response on day 2 were low and only included participants living in non-remote areas, we could not use day 2 intake data to quantify usual nutrient intake. As the vitamin D intake quantified from a single day is unsuitable for assessing nutrient adequacy, we could not compare vitamin D intakes to the estimated average requirement or tolerable upper intake level (47). While the food consumption data used in this study were collected in 2012–2013, this is the first and only nationally representative nutrition-specific survey conducted among Aboriginal and Torres Strait Islander peoples. We did not include dietary supplement intake when quantifying vitamin D intake as detailed vitamin D supplement data was not collected as part of the NATSINPAS (4). The low percentage [<1% (n = 38)] of supplement consumption in the 24 hours prior to the study interview also suggests that supplement intake would not have a significant impact on the quantification of vitamin D intake. Lastly, despite using the Automated Multiple-Pass Method to assist respondents in providing an accurate 24-hour food recall, self-reporting of food consumption remains a potential source of bias, due to the tendency to report dietary behaviours perceived as 'healthy' (45).

We found that vitamin D intake was low among Aboriginal and Torres Strait Islander peoples in Australia. Using findings from this study, along with the vitamin D intake of the general Australian population, food fortification scenarios could be modelled to provide evidence to inform food-based policies to promote vitamin D sufficiency across Australia.

**Acknowledgements**

**Author Contribution Statement:** LJB and CS designed research; BN and AD conducted research and analyzed data; BN wrote the paper; ED, LJB, CS, NN, DT, JJ, BF, CP, CM, and AD reviewed and edited the paper, LJB, ED, CS, NN, DT and JJ provided supervision. LJB had primary responsibility for the final content. All authors have read and approved the final manuscript.

**Data Availability Statement:** Data described in the manuscript were source from publicly available datasets found here: 2012-2013 National Aboriginal and Torres Strait Islander Nutrition and Physical Activity Survey https://www.abs.gov.au/statistics/microdata-tablebuilder/microdatadownload, AUSNUT 2011-2013 https://www.foodstandards.gov.au/science-data/monitoringnutrients/afcd/australian-food-composition-database-download-excel-files#nutrient and vitamin D food composition data https://www.foodstandards.gov.au/science-data/monitoringnutrients/afcd/Data-provided-by-food-companies-and-organisations.

**Funding:** This study was supported by the National Health and Medical Research Council (GNT1184788). BN is supported by a Curtin Strategic Scholarship.

**Competing Interests:** The authors declare no conflict of interest.

**Ethical Approval**: Ethics approval was granted by the Western Australian Aboriginal Health Ethics Committee (WAAHEC), reference HREC979.

**Table 1.** Vitamin D intake of Aboriginal and Torres Strait Islander peoples, stratified by sex and age group.

| Age group (years) | n | n[1] | Bioactivity factor 1 | | | | | | Bioactivity factor 5[2] | | | | | |
|---|---|---|---|---|---|---|---|---|---|---|---|---|---|---|
| | | | mean (µg/d) | SD | 25th | 50th | 75th | 95% CI | mean (µg/d) | SD | 25th | 50th | 75th | 95% CI |
| **Total** | | | | | | | | | | | | | | |
| All ages | 4109 | 606915 | 2.9 | 3.0 | 1.1 | 2.0 | 3.6 | 2.7, 3.0 | 5.3 | 4.3 | 2.4 | 4.2 | 6.8 | 5.1, 5.5 |
| 2 - 3 | 240 | 31011 | 1.9 | 2.2 | 0.8 | 1.2 | 2.1 | 1.5, 2.2 | 3.5 | 2.6 | 2.0 | 3.0 | 4.3 | 3.1, 3.9 |
| 4 - 8 | 515 | 82691 | 2.6 | 2.6 | 1.1 | 1.8 | 3.0 | 2.2, 3.0 | 4.5 | 3.3 | 2.2 | 3.5 | 5.5 | 4.0, 5.0 |
| 9 - 13 | 409 | 70929 | 3.2 | 3.5 | 1.3 | 2.3 | 4.0 | 2.6, 3.7 | 5.3 | 4.1 | 2.9 | 4.6 | 6.6 | 4.7, 5.9 |
| 14 - 18 | 332 | 68823 | 2.7 | 2.4 | 0.9 | 1.9 | 3.8 | 2.3, 3.2 | 5.2 | 3.7 | 2.2 | 4.1 | 7.1 | 4.6, 5.9 |
| 19 - 30 | 722 | 130568 | 3.1 | 2.7 | 1.2 | 2.2 | 4.0 | 2.8, 3.4 | 5.9 | 4.1 | 2.7 | 4.6 | 8.0 | 5.4, 6.4 |
| 31 - 50 | 1098 | 146120 | 3.0 | 3.0 | 1.3 | 2.2 | 3.8 | 2.8, 3.2 | 5.6 | 4.8 | 2.7 | 4.4 | 7.4 | 5.3, 6.0 |
| 51 - 70 | 722 | 70659 | 2.9 | 3.8 | 1.1 | 2.0 | 3.4 | 2.4, 3.3 | 5.1 | 5.2 | 2.3 | 4.1 | 6.7 | 4.6, 5.7 |
| ≥ 71 | 71 | 6114 | 2.1 | 3.4 | 0.8 | 1.0 | 2.4 | 1.3, 2.9 | 3.9 | 5.4 | 2.1 | 2.7 | 4.3 | 2.8, 5.1 |
| **Male** | | | | | | | | | | | | | | |
| All ages | 1814 | 301992 | 3.2 | 3.1 | 1.3 | 2.2 | 4.1 | 3.0, 3.4 | 5.9 | 4.4 | 2.8 | 4.6 | 7.6 | 5.6, 6.2 |
| 2 - 3 | 125 | 16947 | 1.8 | 1.9 | 1.0 | 1.3 | 1.8 | 1.4, 2.3 | 3.5 | 2.3 | 2.2 | 3.1 | 4.3 | 3.0, 4.1 |
| 4 - 8 | 259 | 40953 | 2.8 | 3.1 | 1.1 | 1.8 | 3.6 | 2.1, 3.6 | 4.7 | 3.7 | 2.3 | 3.5 | 6.1 | 3.9, 5.6 |
| 9 - 13 | 207 | 37866 | 3.7 | 4.3 | 1.7 | 2.5 | 4.2 | 2.8, 4.6 | 5.9 | 4.8 | 3.2 | 4.9 | 7.2 | 4.9, 6.9 |
| 14 - 18 | 164 | 34093 | 3.0 | 2.2 | 1.2 | 2.4 | 4.0 | 2.5, 3.5 | 5.9 | 3.7 | 2.7 | 5.0 | 8.0 | 5.0, 6.8 |
| 19 - 30 | 299 | 66534 | 3.5 | 2.6 | 1.4 | 2.4 | 4.8 | 3.0, 4.0 | 6.8 | 4.0 | 3.3 | 5.4 | 9.1 | 6.0, 7.5 |
| 31 - 50 | 427 | 68967 | 3.4 | 2.9 | 1.5 | 2.5 | 4.4 | 3.1, 3.8 | 6.5 | 4.6 | 3.6 | 4.9 | 8.2 | 5.9, 7.1 |
| 51 - 70 | 308 | 34470 | 2.8 | 3.6 | 1.1 | 1.9 | 3.5 | 2.3, 3.2 | 5.3 | 4.9 | 2.7 | 4.5 | 6.8 | 4.7, 6.0 |
| ≥ 71 | 25 | 2162 | 2.9 | 4.0 | 0.9 | 1.6 | 4.0 | 1.3, 4.4 | 4.8 | 4.7 | 2.2 | 3.0 | 5.7 | 3.0, 6.7 |
| **Female** | | | | | | | | | | | | | | |
| All ages | 2295 | 304923 | 2.6 | 2.7 | 1.0 | 1.9 | 3.2 | 2.4, 2.7 | 4.7 | 4.0 | 2.2 | 3.7 | 6.1 | 4.4, 4.9 |
| 2 - 3 | 115 | 14065 | 1.9 | 2.5 | 0.6 | 1.1 | 2.4 | 1.4, 2.3 | 3.5 | 3.0 | 1.8 | 2.7 | 4.1 | 2.9, 4.1 |
| 4 - 8 | 256 | 41737 | 2.3 | 1.9 | 1.0 | 1.8 | 2.8 | 2.0, 2.7 | 4.2 | 2.9 | 2.1 | 3.6 | 5.3 | 3.7, 4.8 |
| 9 - 13 | 202 | 33064 | 2.6 | 1.8 | 1.2 | 2.2 | 3.4 | 2.2, 3.0 | 4.7 | 2.9 | 2.5 | 3.7 | 6.0 | 4.0, 5.3 |
| 14 - 18 | 168 | 34730 | 2.5 | 2.5 | 0.7 | 1.4 | 2.9 | 1.8, 3.2 | 4.6 | 3.6 | 1.7 | 2.9 | 6.4 | 3.6, 5.6 |
| 19 - 30 | 423 | 64034 | 2.7 | 2.7 | 1.0 | 2.0 | 3.4 | 2.4, 3.1 | 5.0 | 3.9 | 2.2 | 4.0 | 6.6 | 4.5, 5.5 |
| 31 - 50 | 671 | 77153 | 2.6 | 3.0 | 1.0 | 1.9 | 3.3 | 2.3, 2.8 | 4.8 | 4.7 | 2.3 | 3.9 | 6.4 | 4.4, 5.2 |
| 51 - 70 | 414 | 36189 | 2.9 | 4.0 | 1.1 | 2.0 | 3.4 | 2.1, 3.7 | 5.0 | 5.4 | 2.1 | 3.8 | 6.7 | 4.1, 5.8 |
| ≥ 71 | 46 | 3951 | 1.7 | 2.9 | 0.8 | 1.0 | 2.3 | 1.0, 2.5 | 3.5 | 5.6 | 2.0 | 2.4 | 3.3 | 2.1, 4.9 |

[1]Weighted to the Aboriginal and Torres Strait Islander population 2012-2013
[2]A bioactivity factor of 5 was used as 25‐hydroxyvitamin D may be up to five times more bioactive than vitamin D
SD, standard deviation; 95% CI, 95% Confidence Interval.

**Table 2.** Vitamin D intake of Aboriginal and Torres Strait Islander peoples, stratified by the remoteness of location and age group.

| Age group (years) | n | n[1] | Bioactivity factor 1 | | | | | | Bioactivity factor 5[2] | | | | | |
|---|---|---|---|---|---|---|---|---|---|---|---|---|---|---|
| | | | mean (µg/d) | SD | 25th | 50th | 75th | 95% CI | mean (µg/d) | SD | 25th | 50th | 75th | 95% CI |
| **Non-remote** | | | | | | | | | | | | | | |
| All ages | 1792 | 477901 | 2.9 | 2.2 | 1.1 | 2.0 | 3.6 | 2.7, 3.1 | 5.2 | 3.2 | 2.4 | 4.2 | 6.7 | 5.0, 5.5 |
| 2 - 3 | 102 | 24762 | 1.9 | 1.4 | 0.9 | 1.3 | 2.1 | 1.5, 2.3 | 3.6 | 1.8 | 2.2 | 3.1 | 4.3 | 3.2, 4.1 |
| 4 - 8 | 214 | 67058 | 2.7 | 1.9 | 1.1 | 1.8 | 3.1 | 2.2, 3.2 | 4.6 | 2.5 | 2.2 | 3.5 | 5.5 | 3.9, 5.2 |
| 9 - 13 | 176 | 56286 | 3.3 | 2.8 | 1.3 | 2.4 | 4.2 | 2.6, 3.9 | 5.3 | 3.2 | 2.9 | 4.5 | 6.5 | 4.6, 6.1 |
| 14 - 18 | 153 | 55473 | 2.7 | 1.8 | 0.9 | 1.9 | 3.8 | 2.2, 3.2 | 5.2 | 2.8 | 2.1 | 4.2 | 7.1 | 4.4, 6.0 |
| 19 - 30 | 332 | 102660 | 3.1 | 2.0 | 1.3 | 2.1 | 4.0 | 2.7, 3.5 | 5.8 | 3.1 | 2.6 | 4.5 | 7.8 | 5.2, 6.4 |
| 31 - 50 | 482 | 112355 | 2.9 | 2.0 | 1.3 | 2.3 | 3.7 | 2.7, 3.2 | 5.5 | 3.4 | 2.7 | 4.5 | 7.3 | 5.1, 5.9 |
| 51 - 70 | 300 | 54231 | 3.0 | 3.0 | 1.1 | 2.0 | 3.5 | 2.4, 3.5 | 5.2 | 3.9 | 2.3 | 4.1 | 6.7 | 4.5, 5.9 |
| ≥ 71 | 33 | 5076 | 2.2 | 2.7 | 0.8 | 1.0 | 2.4 | 1.2, 3.1 | 3.9 | 4.2 | 2.1 | 2.6 | 4.2 | 2.6, 5.3 |
| **Remote** | | | | | | | | | | | | | | |
| All ages | 2317 | 129014 | 2.8 | 4.8 | 1.0 | 2.0 | 3.5 | 2.6, 3.0 | 5.4 | 7.3 | 2.4 | 4.2 | 7.1 | 5.2, 5.7 |
| 2 - 3 | 138 | 6249 | 1.6 | 4.7 | 0.6 | 1.0 | 1.8 | 1.1, 2.1 | 3.1 | 5.4 | 1.6 | 2.6 | 3.6 | 2.5, 3.6 |
| 4 - 8 | 301 | 15633 | 2.2 | 3.3 | 0.9 | 1.6 | 2.6 | 1.9, 2.4 | 4.1 | 5.0 | 2.1 | 3.5 | 5.1 | 3.6, 4.5 |
| 9 - 13 | 233 | 14644 | 2.7 | 3.2 | 1.3 | 2.2 | 3.1 | 2.4, 3.0 | 5.3 | 4.9 | 2.7 | 4.9 | 6.7 | 4.7, 5.8 |
| 14 - 18 | 179 | 13350 | 2.8 | 3.8 | 1.1 | 2.0 | 3.6 | 2.4, 3.3 | 5.5 | 6.2 | 2.5 | 4.1 | 7.1 | 4.7, 6.3 |
| 19 - 30 | 390 | 27909 | 3.3 | 4.6 | 1.2 | 2.5 | 4.2 | 2.9, 3.7 | 6.4 | 7.1 | 2.9 | 5.0 | 8.4 | 5.7, 7.0 |
| 31 - 50 | 616 | 33765 | 3.1 | 6.1 | 1.2 | 2.0 | 3.8 | 2.7, 3.5 | 6.0 | 8.9 | 2.6 | 4.3 | 7.5 | 5.3, 6.7 |
| 51 - 70 | 422 | 16427 | 2.5 | 4.6 | 1.0 | 1.9 | 3.0 | 2.2, 2.8 | 5.0 | 7.3 | 2.3 | 4.1 | 6.9 | 4.5, 5.6 |
| ≥ 71 | 38 | 1037 | 2.0 | 4.8 | 0.6 | 1.2 | 2.4 | 1.3, 2.7 | 4.0 | 6.7 | 1.7 | 3.9 | 4.7 | 2.9, 5.0 |

[1]Weighted to the Aboriginal and Torres Strait Islander population 2012-2013
[2]A bioactivity factor of 5 was used as 25‐hydroxyvitamin D may be up to five times more bioactive than vitamin D
SD, standard deviation; 95% CI, 95% Confidence Interval.

**Table 3.** Top 10 food group contributors to vitamin D intake among Aboriginal and Torres Strait Islander peoples[1]

| Bioactivity factor 1 | % | 95% CI | Bioactivity factor 5[2] | % | 95% CI |
|---|---|---|---|---|---|
| Meat, poultry, game products and dishes | 21.0 | 0.20, 0.22 | Meat, poultry, game products and dishes | 32.7 | 0.32, 0.33 |
| Fats and oils | 19.4 | 0.19, 0.20 | Egg products and dishes | 18.6 | 0.18, 0.19 |
| Egg products and dishes | 14.2 | 0.14, 0.15 | Fats and oils | 10.9 | 0.10, 0.11 |
| Fish and seafood products and dishes | 14.1 | 0.13, 0.15 | Biscuits, pastries, cakes, burgers, and pasta | 9.9 | 0.10, 0.10 |
| Biscuits, pastries, cakes, burgers, and pasta | 10.2 | 0.10, 0.11 | Milk products (e.g., milk, yoghurt, cheese, ice cream) | 9.4 | 0.09, 0.10 |
| Non-alcoholic beverages | 6.9 | 0.06, 0.07 | Fish and seafood products and dishes | 8.0 | 0.08, 0.08 |
| Milk products (e.g., milk, yoghurt, cheese, ice cream) | 5.2 | 0.05, 0.06 | Non-alcoholic beverages | 4.1 | 0.04, 0.04 |
| Confectionery and cereal/nut/fruit/seed bars | 2.5 | 0.02, 0.03 | Breakfast cereals and bread | 1.6 | 0.01, 0.02 |
| Breakfast cereals and bread | 2.2 | 0.02, 0.03 | Confectionery and cereal/nut/fruit/seed bars | 1.4 | 0.01, 0.02 |
| Vegetable products and dishes | 1.0 | 0.01, 0.01 | Vegetable products and dishes | 0.8 | 0.01, 0.01 |

[1]Calculated as (total vitamin D intake from food group/total vitamin D intake from all foods) x 100

[2]A bioactivity factor of 5 was used as 25‐hydroxyvitamin D may be up to five times more bioactive than vitamin D

95% CI, 95% Confidence Interval.

**Supplementary Table 1.** Vitamin D intake of Aboriginal and Torres Strait Islander peoples living in non-remote areas, stratified by sex and age group.

| Age group (years) | n | n[1] | Bioactivity factor 1 | | | | | | Bioactivity factor 5[2] | | | | | |
|---|---|---|---|---|---|---|---|---|---|---|---|---|---|---|
| | | | mean (µg/d) | SD | Percentile 25th | 50th | 75th | 95% CI | mean (µg/d) | SD | Percentile 25th | 50th | 75th | 95% CI |
| **Male** | | | | | | | | | | | | | | |
| All ages | 797 | 237696 | 3.3 | 2.4 | 1.3 | 2.3 | 4.1 | 3.0, 3.5 | 5.9 | 3.3 | 2.9 | 4.6 | 7.6 | 5.5, 6.3 |
| 2 - 3 | 47 | 13161 | 1.9 | 1.4 | 1.1 | 1.3 | 2.0 | 1.3, 2.6 | 3.7 | 1.6 | 2.2 | 3.1 | 4.3 | 3.0, 4.4 |
| 4 - 8 | 106 | 33883 | 2.9 | 2.3 | 1.2 | 1.8 | 3.7 | 2.0, 3.9 | 4.9 | 2.8 | 2.3 | 3.4 | 6.1 | 3.8, 5.9 |
| 9 - 13 | 97 | 29465 | 3.9 | 3.7 | 1.5 | 2.6 | 4.2 | 2.8, 5.1 | 5.9 | 4.0 | 3.2 | 4.6 | 7.2 | 4.7, 7.2 |
| 14 - 18 | 83 | 27698 | 3.0 | 1.7 | 1.2 | 2.5 | 4.0 | 2.4, 3.6 | 5.9 | 2.9 | 2.8 | 5.5 | 8.0 | 4.9, 7.0 |
| 19 - 30 | 137 | 52572 | 3.6 | 2.0 | 1.4 | 2.4 | 4.8 | 3.0, 4.2 | 6.8 | 3.1 | 3.4 | 5.4 | 9.6 | 5.9, 7.8 |
| 31 - 50 | 177 | 52474 | 3.5 | 2.0 | 1.6 | 2.7 | 4.4 | 3.0, 3.9 | 6.5 | 3.2 | 3.8 | 5.1 | 8.2 | 5.8, 7.2 |
| 51 - 70 | 138 | 26721 | 2.9 | 2.9 | 1.2 | 2.0 | 3.5 | 2.3, 3.4 | 5.4 | 3.8 | 2.9 | 4.7 | 6.8 | 4.6, 6.2 |
| $\geq 71$ | 12 | 1722 | 3.1 | 3.4 | 0.8 | 1.6 | 4.0 | 1.1, 5.0 | 4.9 | 4.0 | 2.1 | 2.9 | 5.7 | 2.6, 7.1 |
| **Female** | | | | | | | | | | | | | | |
| All ages | 995 | 240205 | 2.5 | 1.9 | 1.0 | 1.9 | 3.1 | 2.3, 2.7 | 4.6 | 2.9 | 2.1 | 3.6 | 6.0 | 4.3, 4.8 |
| 2 - 3 | 55 | 11602 | 1.9 | 1.5 | 0.7 | 1.4 | 2.5 | 1.4, 2.4 | 3.6 | 1.9 | 2.1 | 3.0 | 5.0 | 2.9, 4.2 |
| 4 - 8 | 108 | 33175 | 2.4 | 1.4 | 1.1 | 1.9 | 2.8 | 2.0, 2.8 | 4.3 | 2.1 | 2.2 | 3.6 | 5.3 | 3.6, 4.9 |
| 9 - 13 | 79 | 26821 | 2.6 | 1.2 | 1.2 | 2.2 | 3.5 | 2.2, 3.0 | 4.7 | 2.0 | 2.5 | 3.9 | 6.1 | 4.0, 5.5 |
| 14 - 18 | 70 | 27775 | 2.4 | 1.9 | 0.7 | 1.3 | 2.8 | 1.5, 3.3 | 4.4 | 2.6 | 1.7 | 2.8 | 6.4 | 3.2, 5.7 |
| 19 - 30 | 195 | 50087 | 2.6 | 1.9 | 1.0 | 1.9 | 3.1 | 2.2, 3.0 | 4.7 | 2.7 | 2.2 | 3.7 | 6.4 | 4.1, 5.2 |
| 31 - 50 | 305 | 59881 | 2.4 | 1.9 | 1.0 | 2.0 | 3.2 | 2.2, 2.7 | 4.7 | 3.4 | 2.2 | 3.8 | 6.3 | 4.2, 5.2 |
| 51 - 70 | 162 | 27510 | 3.0 | 3.1 | 1.0 | 1.9 | 3.5 | 2.0, 4.0 | 4.9 | 3.9 | 2.0 | 3.7 | 6.5 | 3.8, 6.0 |
| $\geq 71$ | 21 | 3354 | 1.7 | 2.1 | 0.8 | 1.0 | 1.8 | 0.8, 2.6 | 3.5 | 4.2 | 2.2 | 2.4 | 3.2 | 1.8, 5.1 |

[1]Weighted to the Aboriginal and Torres Strait Islander population 2012-2013
[2]A bioactivity factor of 5 was used as 25-hydroxyvitamin D may be up to five times more bioactive than vitamin D
SD, standard deviation; 95% CI, 95% Confidence Interval.

**Supplementary Table 2.** Vitamin D intake of Aboriginal and Torres Strait Islander peoples living in remote areas, stratified by sex and age group.

| Age group (years) | n | n[1] | Bioactivity factor 1 | | | | | | Bioactivity factor 5[2] | | | | | |
|---|---|---|---|---|---|---|---|---|---|---|---|---|---|---|
| | | | mean (µg/d) | SD | 25th | 50th | 75th | 95% CI | mean (µg/d) | SD | 25th | 50th | 75th | 95% CI |
| **Male** | | | | | | | | | | | | | | |
| All ages | 1017 | 64296 | 2.9 | 4.3 | 1.2 | 2.1 | 3.6 | 2.6, 3.1 | 5.7 | 7.1 | 2.6 | 4.4 | 7.7 | 5.3, 6.2 |
| 2 - 3 | 78 | 3786 | 1.5 | 2.0 | 0.7 | 1.3 | 1.8 | 1.2, 1.8 | 3.1 | 3.1 | 1.8 | 2.9 | 3.6 | 2.6, 3.6 |
| 4 - 8 | 153 | 7070 | 2.2 | 3.1 | 1.0 | 1.7 | 2.7 | 1.9, 2.6 | 4.2 | 4.9 | 2.3 | 3.9 | 4.9 | 3.5, 4.8 |
| 9 - 13 | 110 | 8401 | 2.8 | 2.7 | 1.7 | 2.2 | 3.3 | 2.4, 3.2 | 5.9 | 4.2 | 3.5 | 6.4 | 7.4 | 5.2, 6.6 |
| 14 - 18 | 81 | 6395 | 3.0 | 3.8 | 1.3 | 2.3 | 4.3 | 2.3, 3.8 | 5.8 | 6.2 | 2.6 | 4.0 | 8.5 | 4.6, 7.0 |
| 19 - 30 | 162 | 13962 | 3.4 | 4.2 | 1.2 | 2.6 | 4.4 | 2.7, 4.0 | 6.5 | 6.5 | 2.9 | 6.1 | 8.0 | 5.6, 7.4 |
| 31 - 50 | 250 | 16494 | 3.3 | 5.2 | 1.3 | 2.1 | 4.3 | 2.7, 3.9 | 6.7 | 8.9 | 3.0 | 4.3 | 8.3 | 5.4, 8.0 |
| 51 - 70 | 170 | 7748 | 2.4 | 4.4 | 1.0 | 1.6 | 3.0 | 2.0, 2.8 | 4.9 | 6.8 | 2.2 | 3.9 | 6.3 | 4.1, 5.7 |
| ≥ 71 | 13 | 440 | 2.1 | 3.3 | 0.9 | 1.3 | 3.4 | 1.2, 3.0 | 4.7 | 4.6 | 3.1 | 4.3 | 4.8 | 3.4, 5.9 |
| **Female** | | | | | | | | | | | | | | |
| All ages | 1300 | 64718 | 2.7 | 5.4 | 1.0 | 1.9 | 3.3 | 2.5, 3.0 | 5.1 | 7.5 | 2.3 | 4.0 | 6.5 | 4.8, 5.5 |
| 2 - 3 | 60 | 2463 | 1.7 | 7.5 | 0.5 | 0.8 | 1.2 | 0.6, 2.8 | 3.1 | 8.1 | 1.6 | 1.8 | 3.2 | 1.9, 4.2 |
| 4 - 8 | 148 | 8562 | 2.1 | 3.4 | 0.7 | 1.5 | 2.6 | 1.7, 2.5 | 4.0 | 4.9 | 1.8 | 3.3 | 5.2 | 3.4, 4.6 |
| 9 - 13 | 123 | 6243 | 2.5 | 3.8 | 1.1 | 2.0 | 3.1 | 2.0, 3.0 | 4.5 | 5.4 | 2.4 | 3.5 | 5.3 | 3.7, 5.2 |
| 14 - 18 | 98 | 6955 | 2.7 | 3.7 | 0.9 | 1.9 | 3.4 | 2.0, 3.3 | 5.2 | 6.0 | 2.4 | 4.2 | 6.3 | 4.0, 6.3 |
| 19 - 30 | 228 | 13947 | 3.2 | 5.0 | 1.2 | 2.4 | 4.0 | 2.7, 3.8 | 6.2 | 7.6 | 2.8 | 4.5 | 8.5 | 5.4, 7.1 |
| 31 - 50 | 366 | 17272 | 3.0 | 6.9 | 1.1 | 1.9 | 3.7 | 2.4, 3.5 | 5.4 | 8.6 | 2.3 | 4.3 | 6.8 | 4.7, 6.0 |
| 51 - 70 | 252 | 8679 | 2.6 | 4.8 | 1.1 | 2.0 | 3.0 | 2.2, 3.0 | 5.1 | 7.7 | 2.5 | 4.3 | 7.1 | 4.5, 5.8 |
| ≥ 71 | 25 | 597 | 1.9 | 5.9 | 0.3 | 0.6 | 2.4 | 0.9, 2.8 | 3.5 | 7.9 | 0.9 | 1.8 | 4.5 | 2.1, 4.8 |

[1]Weighted to the Aboriginal and Torres Strait Islander population 2012-2013
[2]A bioactivity factor of 5 was used as 25-hydroxyvitamin D may be up to five times more bioactive than vitamin D
SD, standard deviation; 95% CI, 95% Confidence Interval.

**Supplementary Table 3.** Top 10 food group contributors to vitamin D intake among Aboriginal and Torres Strait Islander peoples living in non-remote areas[1]

| Bioactivity factor 1 | % | 95% CI | Bioactivity factor 5[2] | % | 95% CI |
|---|---|---|---|---|---|
| Fats and oils | 19.2 | 0.18, 0.20 | Meat, poultry, game products and dishes | 28.2 | 0.27, 0.29 |
| Meat, poultry, game products and dishes | 17.6 | 0.17, 0.19 | Egg products and dishes | 15.7 | 0.15, 0.16 |
| Biscuits, pastries, cakes, burgers, and pasta | 12.9 | 0.12, 0.14 | Biscuits, pastries, cakes, burgers, and pasta | 13.1 | 0.12, 0.14 |
| Egg products and dishes | 11.8 | 0.11, 0.13 | Milk products (e.g., milk, yoghurt, cheese, ice cream) | 11.7 | 0.11, 0.12 |
| Fish and seafood products and dishes | 11.1 | 0.10, 0.12 | Fats and oils | 11.1 | 0.10, 0.12 |
| Non-alcoholic beverages | 8.4 | 0.08, 0.09 | Fish and seafood products and dishes | 6.5 | 0.06, 0.07 |
| Milk products (e.g., milk, yoghurt, cheese, ice cream) | 6.2 | 0.06, 0.07 | Non-alcoholic beverages | 5.1 | 0.05, 0.06 |
| Confectionery and cereal/nut/fruit/seed bars | 4.1 | 0.04, 0.05 | Breakfast cereals and bread | 2.4 | 0.02, 0.03 |
| Breakfast cereals and bread | 3.7 | 0.03, 0.04 | Confectionery and cereal/nut/fruit/seed bars | 2.3 | 0.02, 0.03 |
| Meal replacement and protein powders | 1.5 | 0.01, 0.02 | Vegetable dishes | 1.3 | 0.01, 0.02 |

[1]Calculated as (total vitamin D intake from food group/total vitamin D intake from all foods) x 100
[2]A bioactivity factor of 5 was used as 25-hydroxyvitamin D may be up to five times more bioactive than vitamin D
95% CI, 95% Confidence Interval.

**Supplementary Table 4.** Top 10 food group contributors to vitamin D intake among Aboriginal and Torres Strait Islander peoples living in remote areas[1]

| Bioactivity factor 1 | % | 95% CI | Bioactivity factor 5[2] | % | 95% CI |
|---|---|---|---|---|---|
| Meat, poultry, game products and dishes | 23.6 | 0.23, 0.25 | Meat, poultry, game products and dishes | 36.1 | 0.35, 0.37 |
| Fats and oils | 19.5 | 0.19, 0.20 | Egg products and dishes | 20.7 | 0.20, 0.21 |
| Fish and seafood products and dishes | 16.4 | 0.16, 0.17 | Fats and oils | 10.7 | 0.10, 0.11 |
| Egg products and dishes | 16.2 | 0.15, 0.17 | Fish and seafood products and dishes | 9.1 | 0.09, 0.10 |
| Biscuits, pastries, cakes, burgers, and pasta | 8.1 | 0.07, 0.09 | Milk products (e.g., milk, yoghurt, cheese, ice cream) | 7.6 | 0.07, 0.08 |
| Non-alcoholic beverages | 5.8 | 0.05, 0.06 | Biscuits, pastries, cakes, burgers, and pasta | 7.5 | 0.07, 0.08 |
| Milk products (e.g., milk, yoghurt, cheese, ice cream) | 4.3 | 0.04, 0.05 | Non-alcoholic beverages | 3.3 | 0.03, 0.04 |
| Confectionery and cereal/nut/fruit/seed bars | 1.4 | 0.01, 0.02 | Breakfast cereals and bread | 1.0 | 0.01, 0.01 |
| Breakfast cereals and bread | 1.1 | 0.01, 0.01 | Reptiles, amphibia and insects | 0.8 | 0.01, 0.01 |
| Reptiles, amphibia and insects | 1.0 | 0.01, 0.01 | Confectionery and cereal/nut/fruit/seed bars | 0.7 | 0.01, 0.01 |

[1]Calculated as (total vitamin D intake from food group/total vitamin D intake from all foods) x 100
[2]A bioactivity factor of 5 was used as 25-hydroxyvitamin D may be up to five times more bioactive than vitamin D
95% CI, 95% Confidence Interval.